\documentclass[
    ,final            
  ]
  {aipproc}

\layoutstyle{6x9}

\usepackage{xspace}

\newcommand{\pinot}{\ensuremath{\pi^0}\xspace}
\newcommand{\pp}{pp\xspace}
\newcommand{\pbpb}{Pb-Pb\xspace}
\newcommand{\pt}{\ensuremath{p_{T}}\xspace}

\newcommand{\etal}{\emph{et al.}\xspace}

\begin{document}

\title{Jets and Photons in ALICE}


\author{Thomas Dietel for the ALICE Collaboration}{
  address={Westf\"alische-Wilhelms-Universit\"at M\"unster\\
  Wilhelm-Klemm-Str 9\\
  48149 M\"unster\\
  Germany}
}

\begin{abstract}

ALICE measured transverse momentum spectra of $\pi^0$ and $\eta$ mesons via the
two photon decay in pp collisions at $\sqrt{s}=0.9$, 2.76 and 7\,TeV and Pb-Pb
collisions at $\sqrt{s_{NN}}=2.76\,\mbox{TeV}$. NLO pQCD calculations agree with
\pp measurements at 0.9\,TeV, but overestimate the data at 2.76 and 7\,TeV. The
nuclear modification factor for neutral pions shows a strong suppression of
high-\pt particle production in central \pbpb collisions.

Raw spectra of charged particle jets have been measured in \pbpb collisions.
Detailed studies of background fluctuations have been performed and will allow
us to unfold the spectra even for low momentum cut offs, giving access to
soft fragmentation products in quenched jets.

\end{abstract}

\maketitle


\section{Introduction}

The interaction of high-momentum partons with hot and dense nuclear matter is a
key point to the understanding of heavy-ion collisions. The energy loss of these
partons during the traversal of the medium is called jet quenching
\cite{Bjorken:1982}. It results in a suppression of particle production at high
transverse momenta, as measured with the nuclear modification factor $R_{AA}$
for hadrons \cite{STAR:2002:Raa}. A second consequence is the increased
production of soft particles associated with the original high-momentum parton
to ensure energy and momentum conservation. This low-\pt enhancement is
observable in the modification of fragmentation functions derived from fully
reconstructed jets \cite{SalgadoWiedemann:2003}.

\section{ALICE Detector and Data Sets}

The ALICE detector at the CERN LHC was designed for the study of the quark-gluon
plasma created in high-multiplicity heavy-ion collisions. The central
barrel with a large Time Projection Chamber (TPC) and the Inner Tracking System
(ITS) with 6 layers of silicon detectors provides tracking with full
athimuthal acceptance in $|\eta|<0.9$. ALICE also features two electromagnetic
calorimeters at mid-rapidity: the fine-granularity PHOS covering $|\eta|<0.13$
and $\Delta\phi=60^\circ$ and the EMCal with coarser granularity but larger
acceptance of $|\eta|<0.7$ and $\Delta\phi=40^\circ$ in 2010, increased to
$\Delta\phi=100^\circ$ in 2011.

The data presented here are from \pp runs at $\sqrt{s}=0.9$ and $7\,\mbox{TeV}$
taken in 2010, the first \pbpb run at $\sqrt{s_{NN}}=2.76\,\mbox{TeV}$ in fall
2010 and a \pp run at the reference energy of $\sqrt{s}=2.76\,\mbox{TeV}$ in
spring 2011.

\section{Neutral Meson Production in \pp Collisions}

\pt spectra of neutral mesons in \pp{} collisions are not only an important
reference for \pbpb{} collisions, but also test perturbative QCD at the highest
available energies. With calculable parton cross-sections and parton
distribution functions in the relevant kinematic region well under control, the
main uncertainty comes from fragmentation functions. Pion production with
$\pt<20\,\mbox{GeV}/c$ at top LHC energy is dominated by gluon fragmentation,
but the gluon fragmentation functions are poorly constrained by measurements of
\pp collisions at lower energies and $e^+e^-$ collisions. Neutral meson spectra
at LHC energies therefore provide important constraints on gluon fragmentation
functions.

ALICE has measured neutral pions and $\eta$ mesons in \pp collisions at
$\sqrt{s}=0.9$, 2.76 and 7\,TeV using the invariant mass of their two-photon
decays, where the photons were either detected by the PHOS or via conversions
with the TPC and ITS \cite{Reygers:QM2011}.

\begin{figure}
\includegraphics[width=0.48\textwidth]{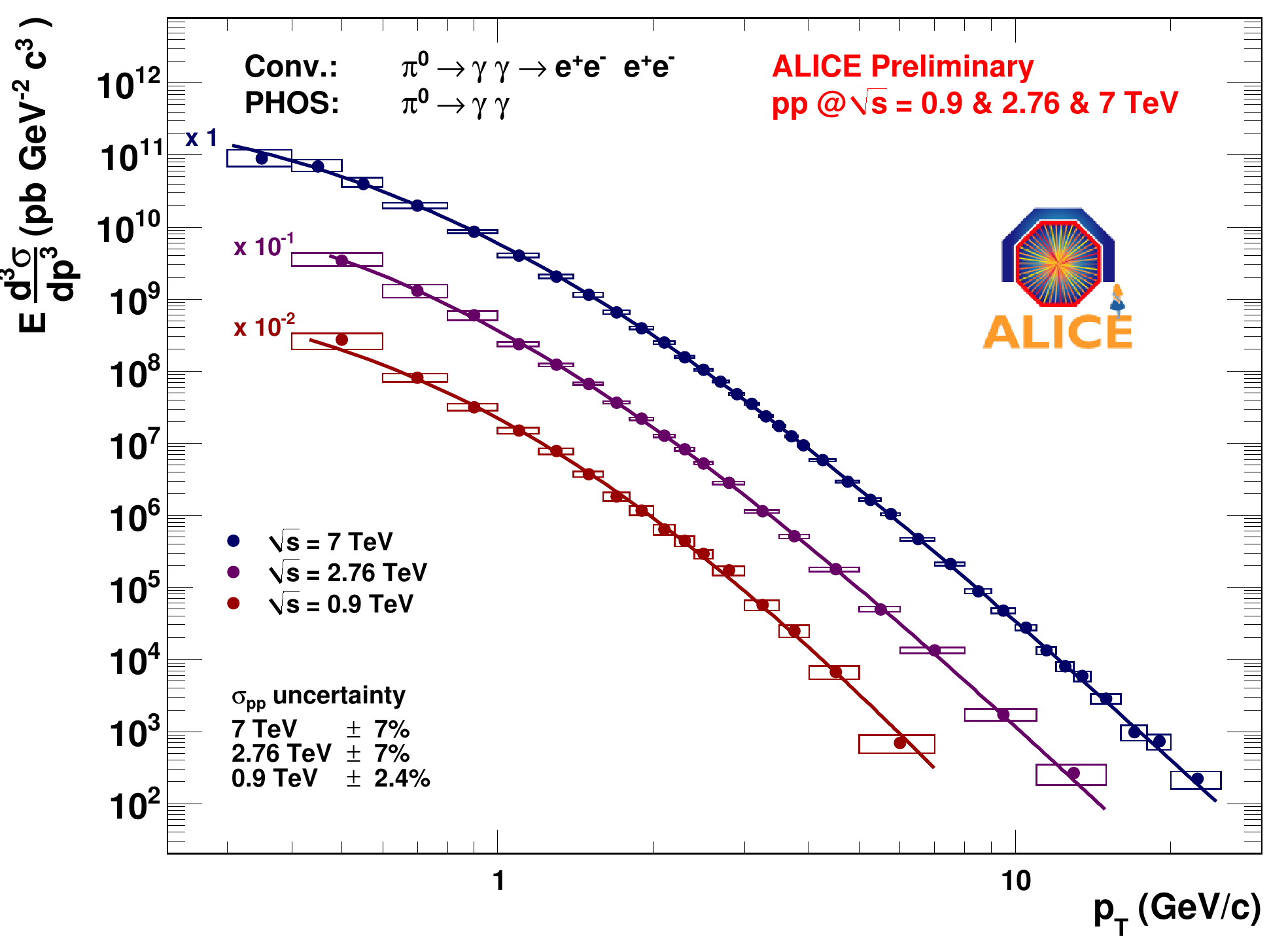}
\includegraphics[width=0.48\textwidth]{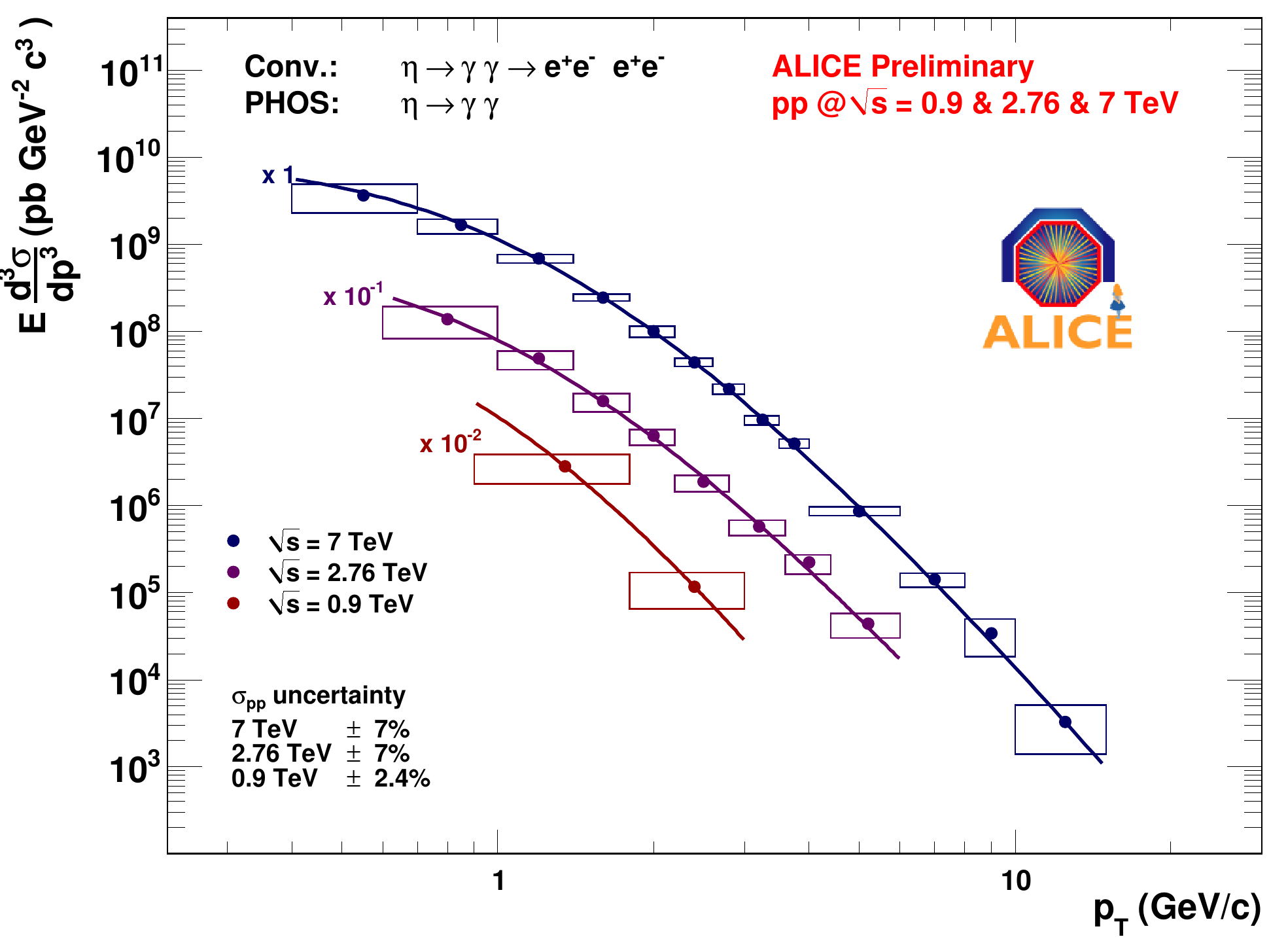}
\caption{\label{fig:eta_pi_xsect}
Invariant cross sections for \pinot(left) and $\eta$ (right) mesons in
\pp collisions at $\sqrt{s}=0.9$, 2.76 and 7\,TeV. The data points represent the weighted average
of conversion and PHOS measurements and were parameterized with a Tsallis
function $E\mbox{d}^3\sigma/\mbox{d}p^3 \propto (1+(m_T - m)/(nT))-n$}
\end{figure}

\begin{figure}
\centering
\includegraphics[width=0.48\textwidth]{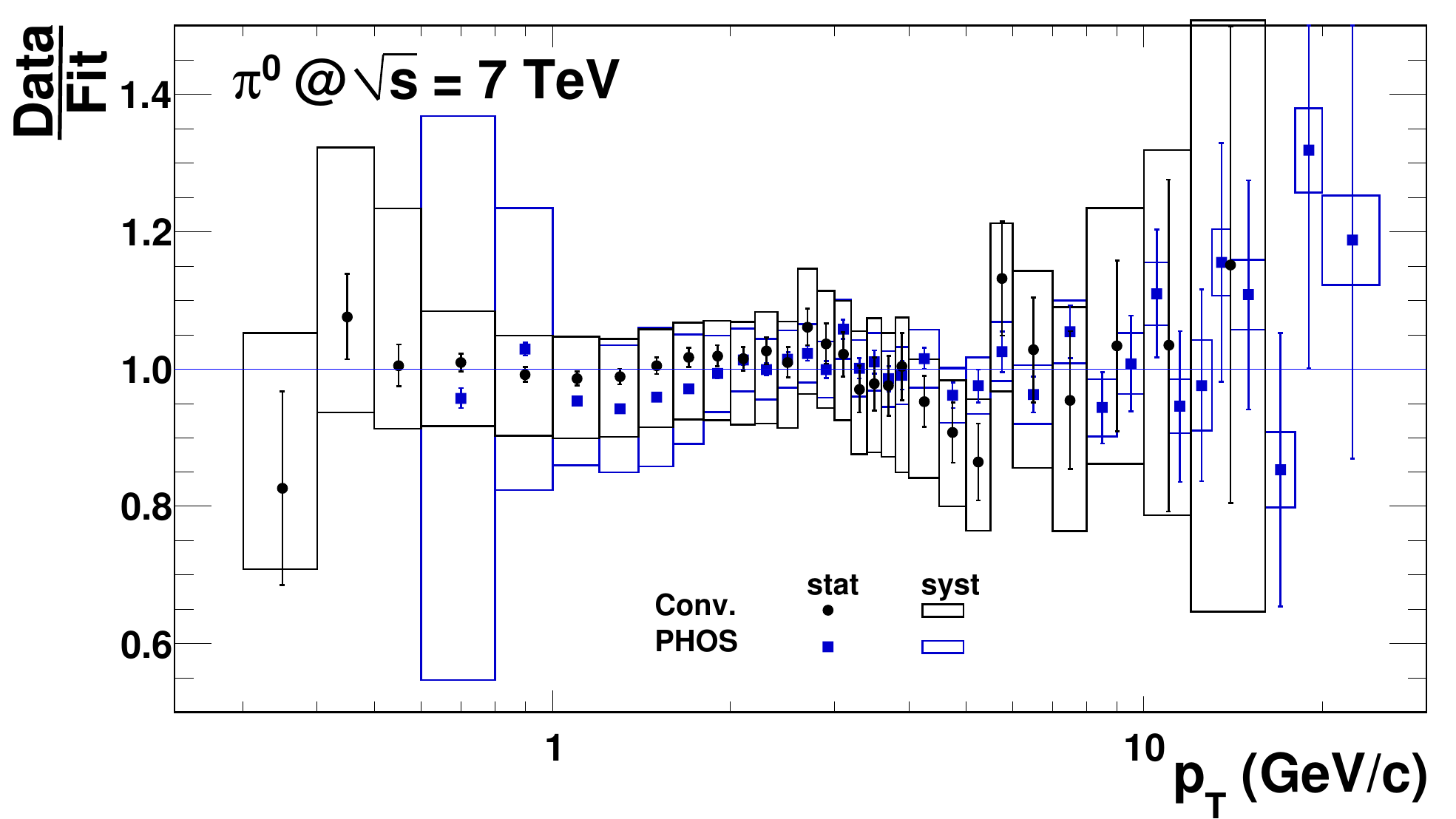}
\includegraphics[width=0.48\textwidth]{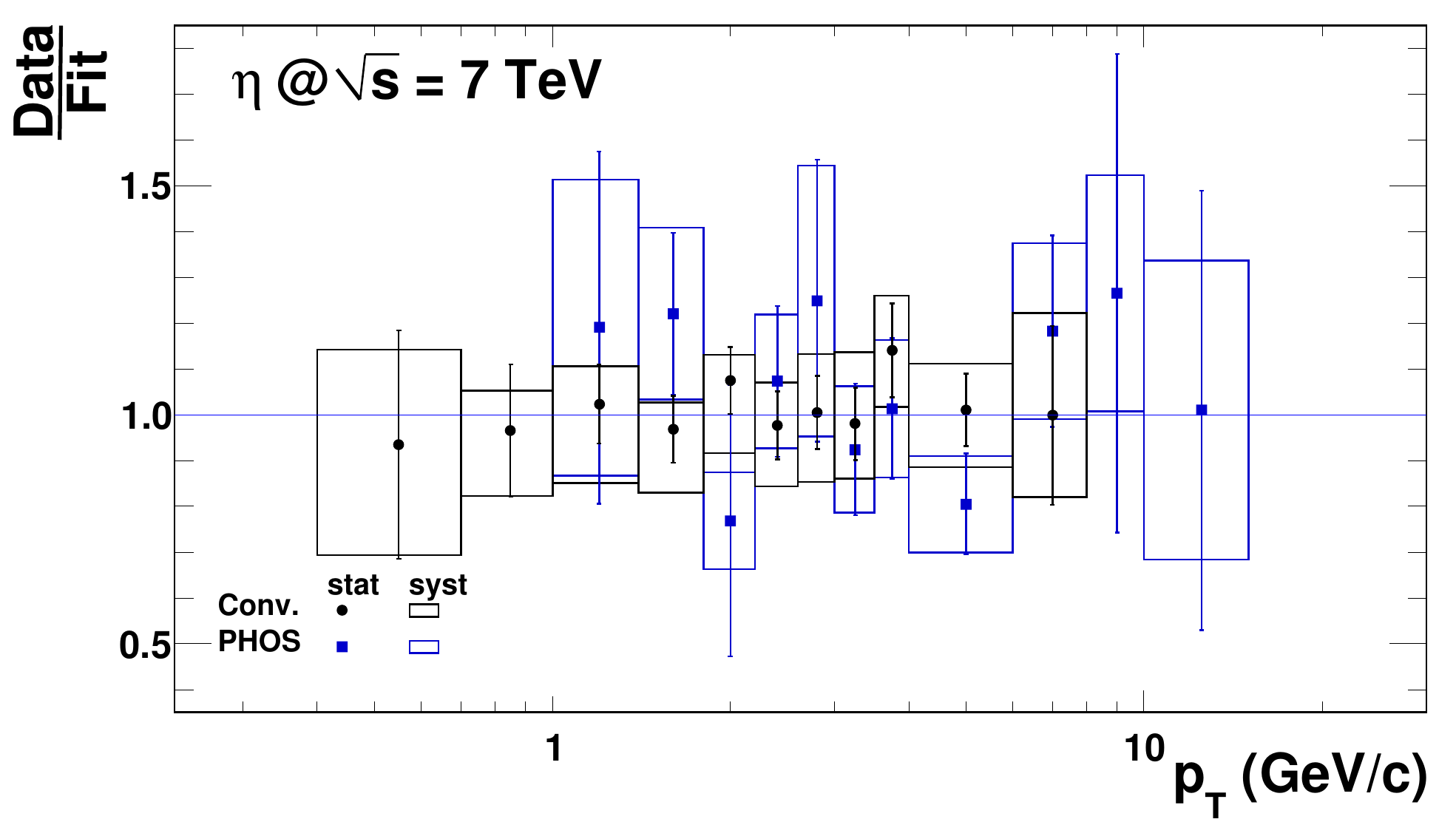}
\caption{\label{fig:eta_pi_conv_phos_ratio}
Ratio of \pt spectra measured with conversions (black) and PHOS (black) to the
Tsallis parameterisation for \pinot(left) and $\eta$ (right) mesons in \pp collisions at 
$\sqrt{s}=7\,\mbox{TeV}$.}
\end{figure}

Figure \ref{fig:eta_pi_xsect} shows the combined invariant cross sections for
$p+p\rightarrow \pinot+X$ (left) and $p+p\rightarrow \eta+X$ (right) obtained
from the conversion and the PHOS measurements of neutral pions and eta mesons.
The data are fully corrected, including a small correction (up to 20\%) to
account for the variation within one bin. The spectra are fitted with a Tsallis
function to facilitate comparisons between the two reconstruction methods and
with NLO calculations.

Figure \ref{fig:eta_pi_conv_phos_ratio} compares the conversion and PHOS methods
to measure the invariant cross section. Shown is the ratio of both methods to
the Tsallis parameterisation of the combined spectra from Figure
\ref{fig:eta_pi_xsect}. Both measurements agree very well within their error
bars.

\begin{figure}
\centering
\includegraphics[width=0.45\textwidth]{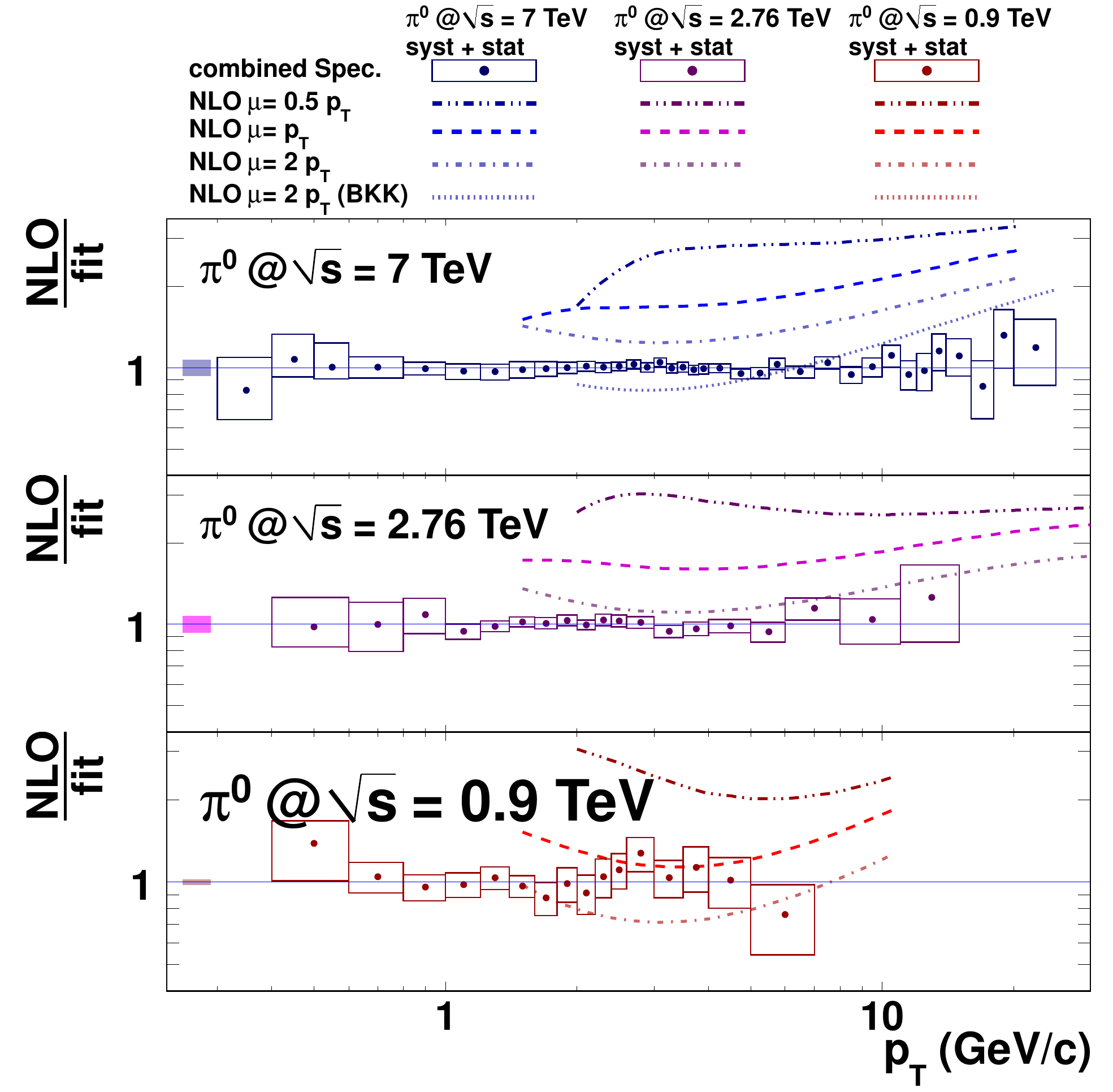}
\includegraphics[width=0.45\textwidth]{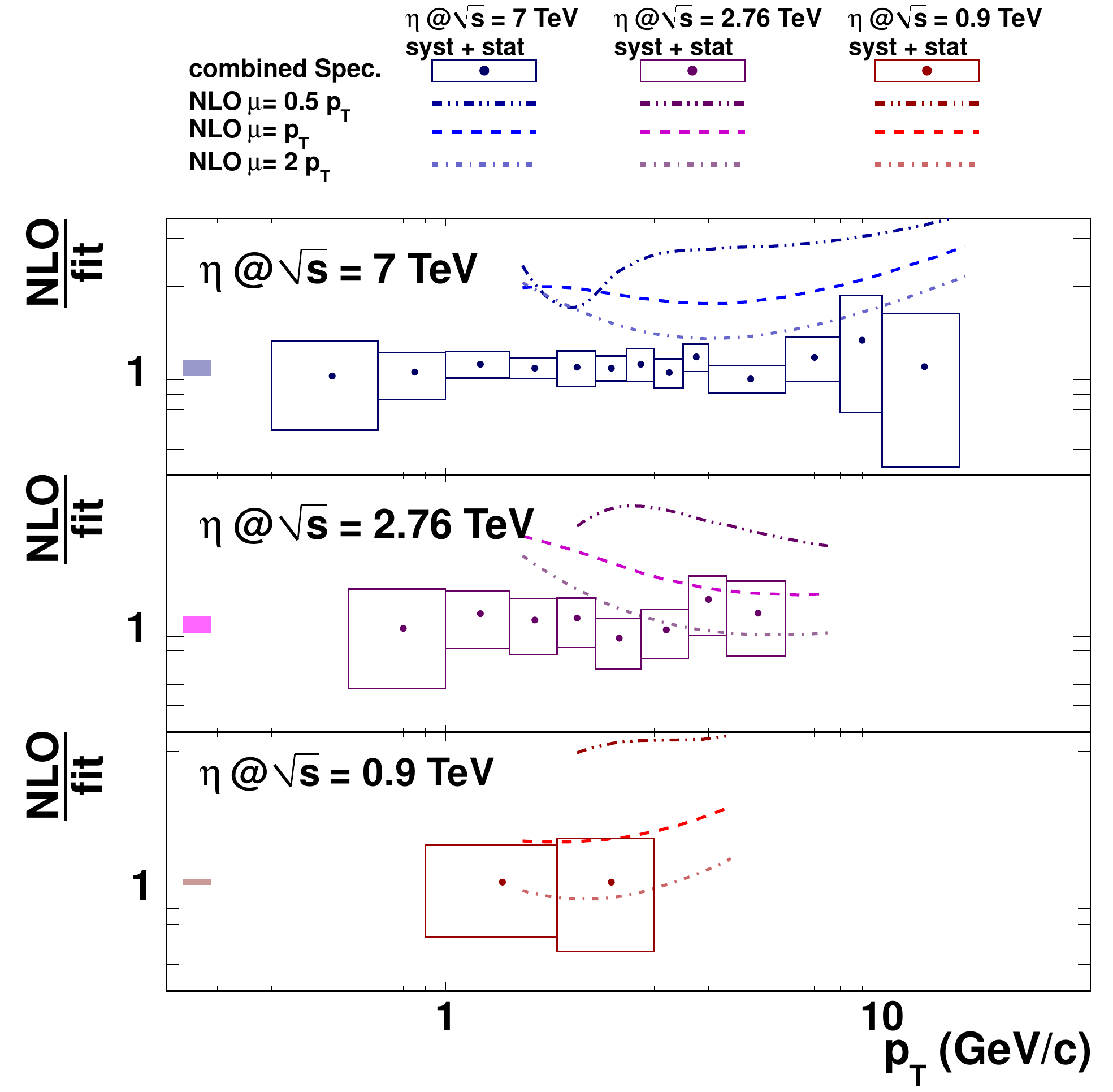}
\caption{\label{fig:eta_pi_nlo}
Comparison of data and NLO pQCD calculations, normalized to the Tsallis fit from
Figure~\ref{fig:eta_pi_xsect}. For details about the calculations see text.}
\end{figure}

The \pinot cross section is compared to an NLO pQCD calculation using the
CTEQ6M5 parton distribution functions \cite{pdf:cteq6m5} and DSS fragmentation
functions \cite{ff:dss}. The theoretical uncertainty of the calculation has been
estimated by varying unphysical scales $\mu=0.5\pt,\pt,2\pt$ \cite{Jager2003}.
Figure \ref{fig:eta_pi_nlo} shows the NLO calculation, normalized to the Tsallis
fit, in comparison with the measured spectra. For $\sqrt{s}=0.9\,\mbox{TeV}$,
the calculation agrees with the data, while for higher energies it overestimates
the data. At 7\,TeV a second calculation \cite{Aurenche1999} using the BKK FF
\cite{ff:bkk} shows better agreeement with the data. The left panel in Figure
\ref{fig:eta_pi_nlo} compares the measured $\eta$ spectra with an NLO
calculation using the AESSS FF \cite{ff:aesss}. The trend is similar to the DSS
FF in pion data, showing agreement at 0.9\,TeV and an overestimation at higher
energies of 2.76 and 7\,TeV.

\section{Nuclear Modification Factor for Pions}

The modification of spectra in heavy-ion collisions is commonly expressed with
the nuclear modification factor $R_{AA}$, defined as the ratio of production
rates in nucleus-nucleus and \pp collisions, scaled by the number of binary
nucleon-nucleon interactions as determined from a Glauber calculation:
\[
R_{AA} = \frac{1}{\langle N_{coll}
\rangle}\frac{(1/N_{evt}^{AA})d^2N_{\pinot}^{AA}/dyd\pt}{(1/N_{evt}^{pp})d^2N_{\pinot}^{pp}/dyd\pt}
\]

\begin{figure}
\centering
\includegraphics[width=0.49\textwidth]{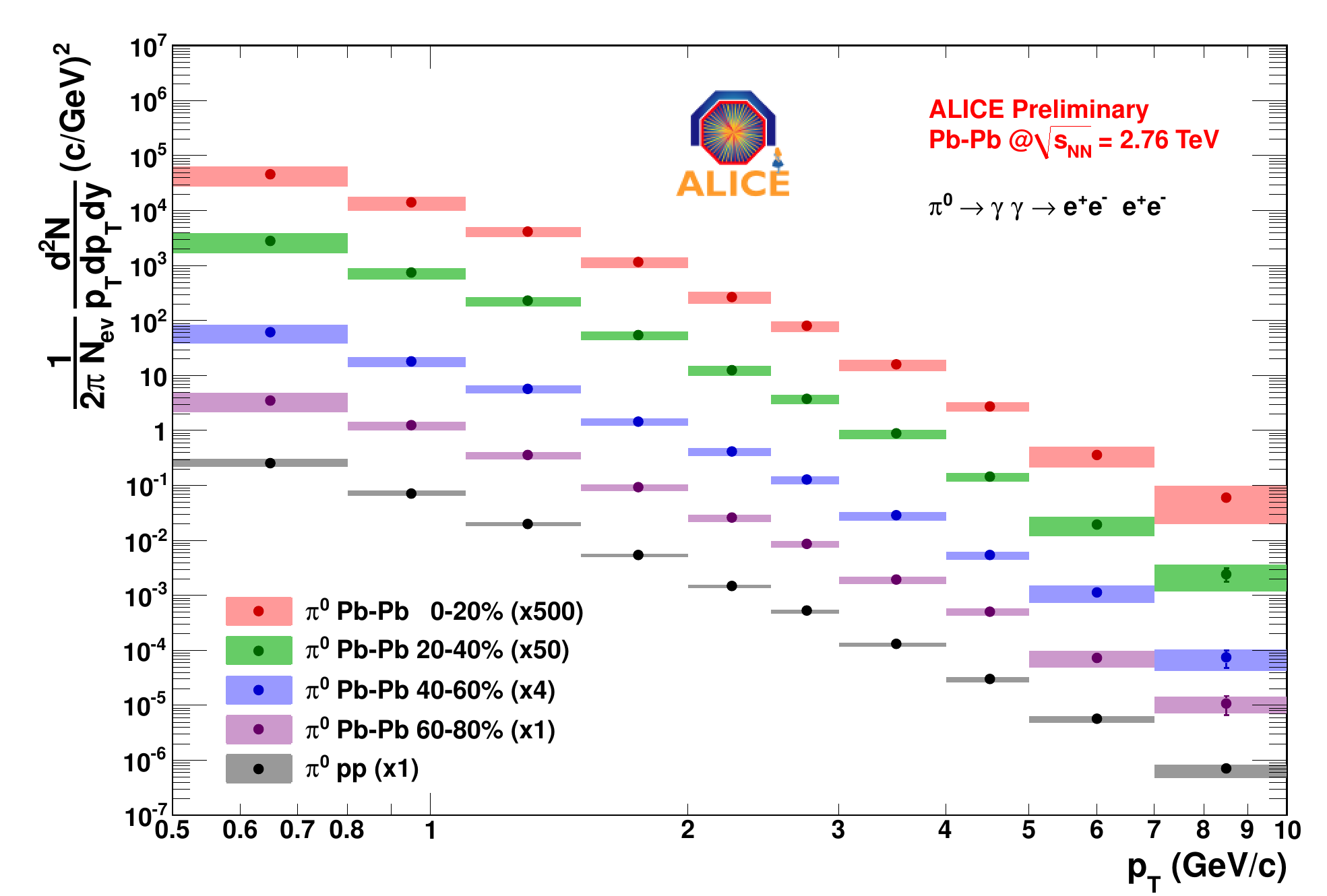}
\includegraphics[width=0.49\textwidth]{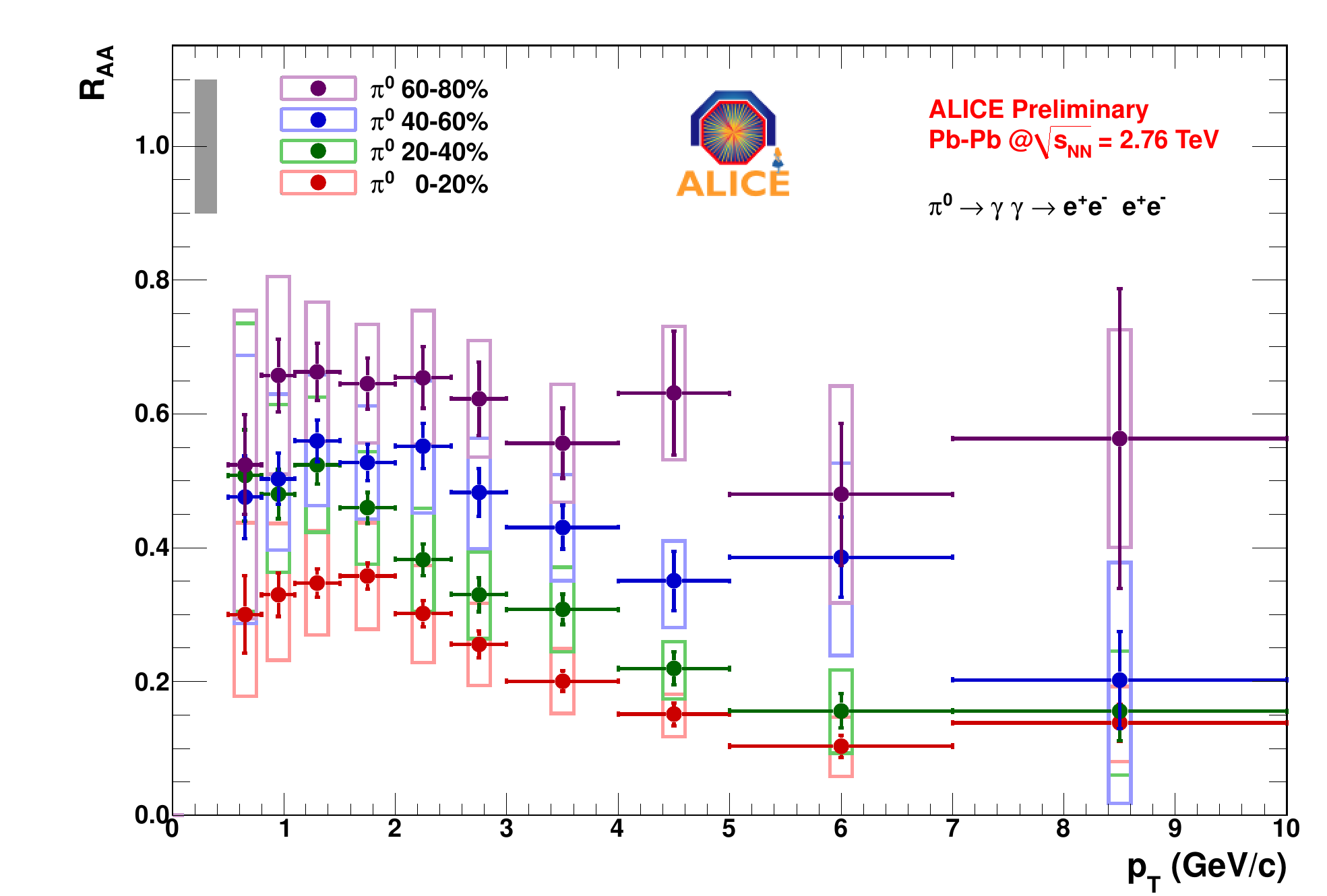}
\caption{\label{fig:pi_raa}
\pinot yield (left) and nuclear modification factor $R_{AA}$ (right) in \pbpb
collisions at $\sqrt{s_NN}=2.76\,\mbox{TeV}$ in 4 centrality bins. The left
panel also shows the \pinot yield in \pp collisions.}
\end{figure}

The yield of neutral pions in \pbpb collisions at $\sqrt{s_{NN}} =
2.76\,\mbox{TeV}$ taken in fall 2010 has been measured by ALICE using the
conversion method \cite{Conesa:QM2011}. The yields as a function of \pt for 4
different centrality bins are shown in the left panel of Figure
\ref{fig:pi_raa}, together with the yield for \pp collisions. The right panel
shows the nuclear modification factor $R_{AA}$ for the same 4 centrality bins. A
clear suppression of high-\pt hadron production can be seen that increases from
peripheral to central collisions. The maximum suppression is reached at $\pt
\approx 6\,\mbox{GeV}/c$ with an $R_{AA}\approx 0.1$, consistent with the
suppression of charged hadrons \cite{alice:2010:raa}. The observed suppression
is stronger than at RHIC, where PHENIX reported a value of 0.2 to 0.3 
\cite{phenix:2003:raa}.

\section{Jet Reconstruction and Background Fluctuations}

Full jet reconstruction can provide a complementary picture to the suppression
of single particles as discussed in the previous section. By accounting for
low-\pt particles, the energy lost due to jet quenching can be recovered, so
that the redistribution of energy and momentum in the quenching process can be
studied in terms of medium-modified fragmentation functions. In order to be
sensitive to jet modifications it is extremely important to minimize the bias on
hard fragmentation by applying the lowest possible momentum cut off during jet
reconstruction.


We applied various cone and sequential recombination jet reconstruction
algorithms to \pbpb collisions. Lacking large acceptance calorimetry in 2010,
the algorithms from the FastJet package were run on charged tracks only with a
radius/distance parameter of 0.4. For this study, we use the anti-$k_T$ jet
finder and subtract the background density $\rho$ as determined by calculating
the median \pt/area of reconstructed $k_T$ clusters \cite{KleinBoesing:QM2011}.

\begin{figure}
\centering
\includegraphics[width=0.46\textwidth]{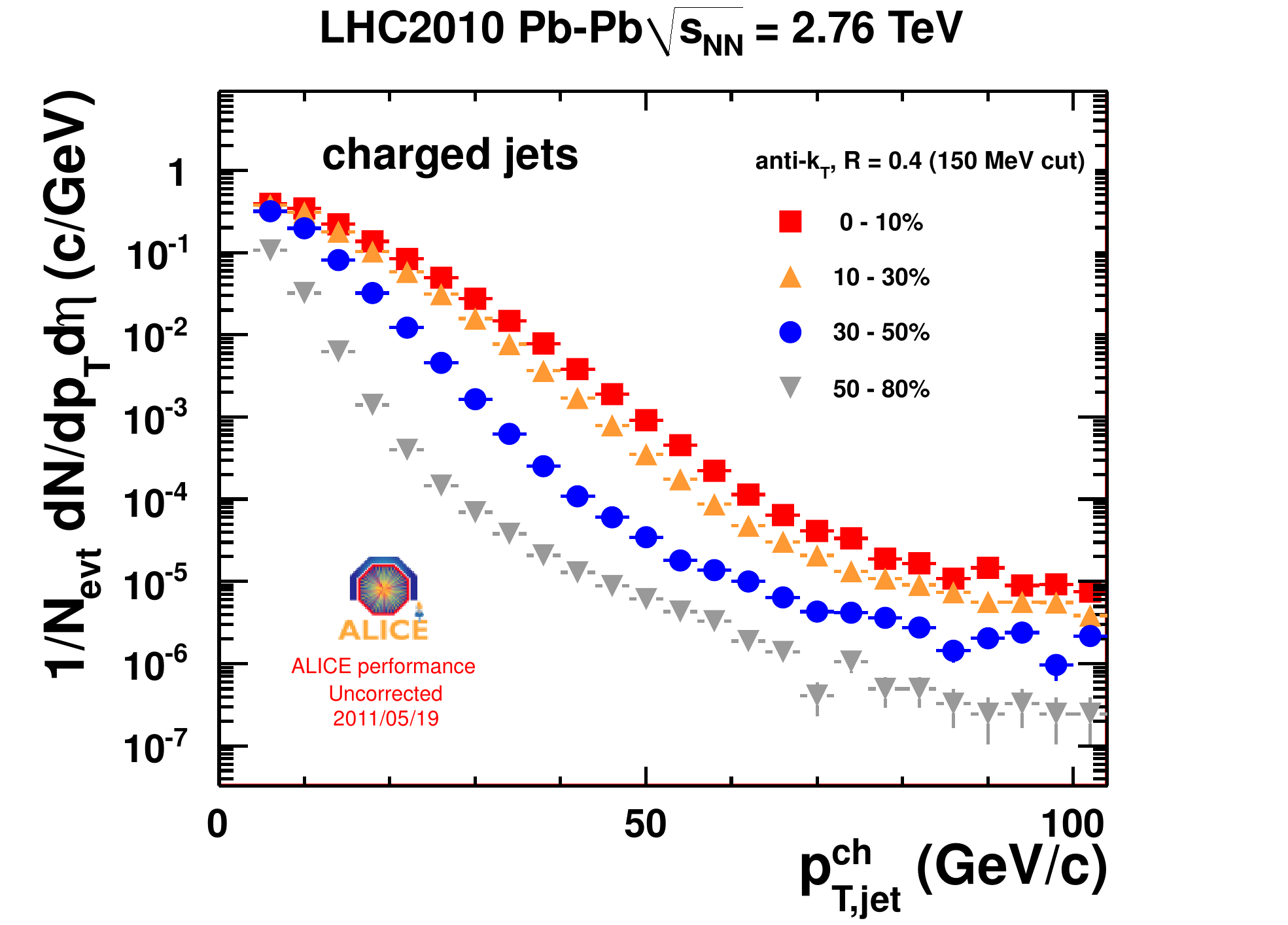}
\hspace{5mm}
\includegraphics[width=0.46\textwidth]{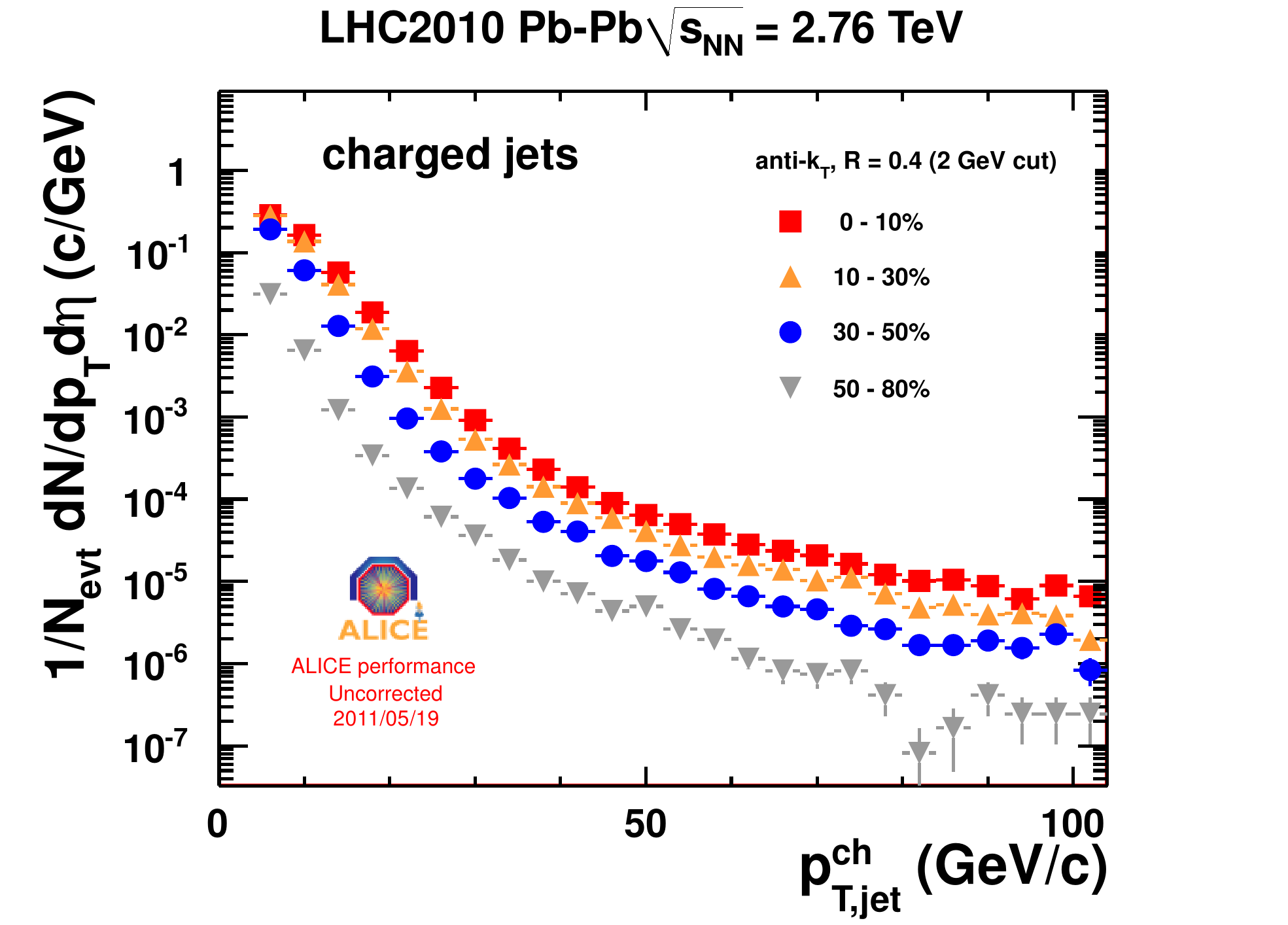}
\caption{\label{fig:jet_spectra_raw}
Raw jet spectra for different centralities using the anti-$k_T$ jet finder on
charged tracks, after background subtraction. In the left panel, a track cut of
$\pt>150\,\mbox{MeV}/c$ was used, in the right panel $\pt>2\,\mbox{GeV}/c$.}
\end{figure}

Figure \ref{fig:jet_spectra_raw} shows raw jet spectra for different
centralities, reconstructed with a track momentum cut off of
$\pt>150\,\mbox{MeV}/c$ (left) and $\pt>2\,\mbox{GeV}/c$ (right). While the
shapes in the peripheral and central event classes are similar for the high
threshold, a strong modification is visible for the low threshold. This
difference is caused by background fluctuations in central events that
predominantly appear at low transverse momenta.

These background fluctuations can be studied using different probes in \pbpb
collisions: random cones (R=0.4), where the \pt of all tracks in the cone is
summed, embedded single high-\pt tracks and embedded PYTHIA-jets. In case of
embedding, the same anti-$k_T$ jet finder as for unmodified events is run and
the reconstructed jets are matched to the embedded probe by finding the track or
50\% of the embedded jet momentum. We then define the residuals of the jet
measurements as \[ \delta \pt = \pt^{rec} - A\cdot \rho - \pt^{probe} \] where
$\pt^{probe}=0$ for random cones. The residuals for random cones are also
analyzed for events where all tracks were randomized in $\eta$ and $\phi$ and
where the leading two jets in each event were excluded.

\begin{figure}
\centering
\includegraphics[width=0.42\textwidth]{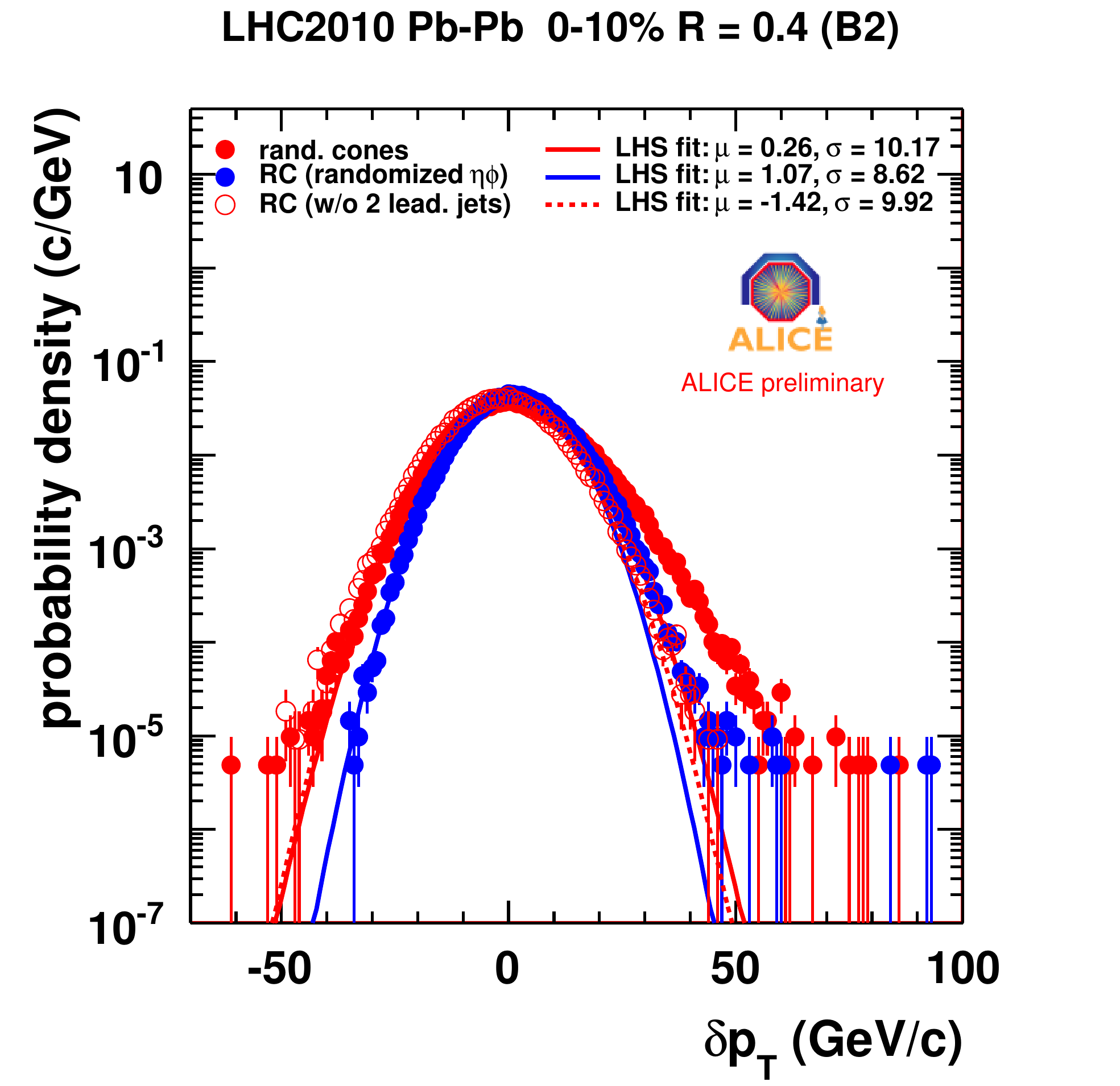}
\hspace{5mm}
\includegraphics[width=0.42\textwidth]{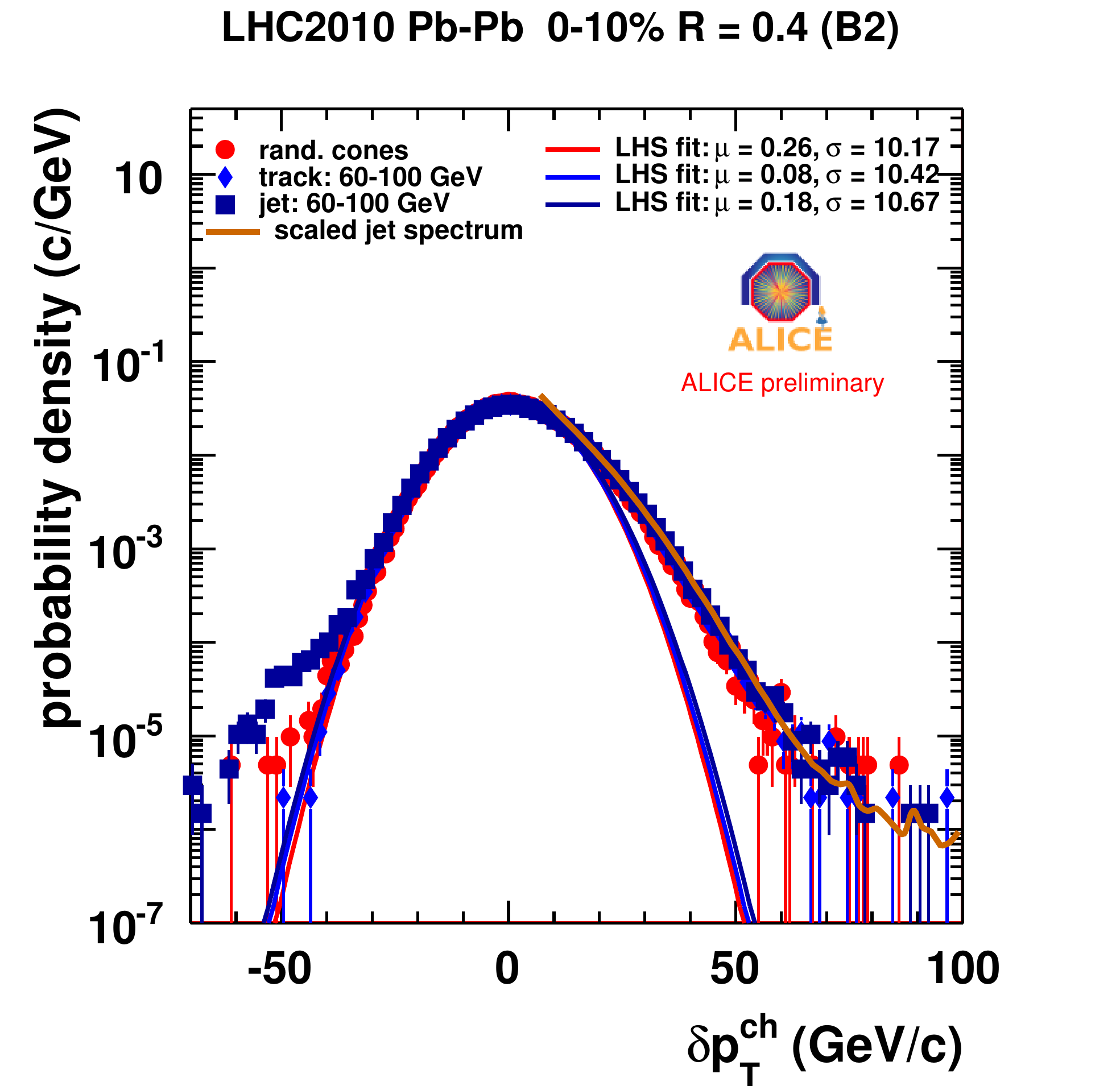}
\caption{\label{fig:jets_dpt}
Left panel: difference of embedded and reconstructed energy (residuals) for
random cones in unmodified and randomized events, and events with the two leading anti-$k_T$
cluster excluded. Right panel: residuals for random cones, embedded tracks and
embedded jets, compared to the scaled raw jet spectrum.}
\end{figure}

The residuals for random cones, shown in the left panel of Figure
\ref{fig:jets_dpt}, exhibit an assymetric shape that follows a Gaussian
distribution for $\delta\pt<0$ and has a tail towards high \pt. This tail can be
attributed to the onset of jet production, which is confirmed when the leading
two jets are excluded: the distribution becomes Gaussian and is only sensitive
to background fluctuations. For randomized events, we again see a Gaussian
distribution, but with a smaller width than before. This indicates that the
randomization did not only destroy jet-like correlations, but also other
region-to-region fluctuations like e.g. elliptic flow. 

The right panel in Figure \ref{fig:jets_dpt} compares the random cones with
embedded tracks and jets. The residuals obtained for these probes are consistent
with each other and also match the scaled raw jet spectrum. This confirms that
the broadening of the jet spectrum for the low momentum cut off is caused by
background fluctuations. Unfolding of the jet spectra will therefore be
necessary. The fact that the mean of the Gaussian distributions, as determined
from fits to the left hand side, is close to 0 affirms the validity of the
background subtraction.

\section{Conclusions}

\pinot and $\eta$ invariant cross sections were measured in \pp collisions at
$\sqrt{s}=0.9$, 2.76 and 7\,TeV. None of the considered NLO pQCD calculations
can reproduce the data at all available energies.

The nuclear modification factor for \pinot production was determined from \pp
and \pbpb collisions at $\sqrt{s_{NN}}=2.76\,\mbox{TeV}$. A stronger suppression
of high-\pt particles than at RHIC energies was found, consistent with charged
particle measurements of ALICE.

Jet reconstruction even in the most central \pbpb collisions is possible, but
subject to large background fluctuations. These fluctuations were studied by
placing random cones and embedded tracks and jets in real events.





\bibliographystyle{aipproc}   



\IfFileExists{\jobname.bbl}{}
 {\typeout{}
  \typeout{******************************************}
  \typeout{** Please run "bibtex \jobname" to optain}
  \typeout{** the bibliography and then re-run LaTeX}
  \typeout{** twice to fix the references!}
  \typeout{******************************************}
  \typeout{}
 }


\end{document}